\documentclass[%
reprint,
nosuperscriptaddress,
nofootinbib,
 amsmath,amssymb,
 aps,
 prd,
 twocolumn,
 floatfix
]{revtex4-2}
\usepackage[colorlinks=true,linkcolor=blue,urlcolor=blue,filecolor=black,citecolor=red,pdfstartview=FitV,pdftitle={},pdfsubject={},pdfkeywords={},pdfpagemode=None,bookmarksopen=true]{hyperref}
\usepackage{graphicx}
\usepackage{amsmath}
\usepackage{amssymb,ulem}
\usepackage{color}%
\usepackage{tikz}
\usepackage{dcolumn}
\usepackage{bbold}
\usepackage{times}
\usepackage{pdfsync}
\usepackage{xcolor}
\usepackage{setspace}
\usepackage[T1]{fontenc}

\newcommand{\SOUTHCUT}{\vspace{0.5em} School of Physics and Optoelectronics, South China University of Technology, Guangzhou 510641, People's Republic of China}
\begin{document}
\title{Black hole spacetimes with dark matter spikes: Energy-momentum tensor and backreaction effects}
\author{Wei Xiong}
\author{Peng-Cheng Li}
\email{pchli2021@scut.edu.cn}
\affiliation{\SOUTHCUT}
\date{\today}
\begin{abstract}
  We study the energy-momentum tensor of a dark matter (DM) spike formed during the adiabatic growth of a black hole embedded in a DM halo, and investigate its backreaction on the spacetime geometry. Within the Einstein cluster framework, we derive the complete tensor, explicitly incorporating the kinetic contribution to the energy density and the anisotropic pressure arising from noncircular particle orbits. Adopting the Hernquist profile as an illustrative model of DM halo and employing parameters appropriate to the Milky Way, we find that near the spike, the kinetic term enhances the total energy density by approximately 50\% relative to the rest-mass component, while the nonzero radial pressure induces a mild anisotropy in the stress tensor. The derived tensor satisfies all standard energy conditions. By treating it as a fixed source in Einstein’s equations, we numerically obtain a static, spherically symmetric metric that deviates from the Schwarzschild solution by an amount more than twice that found when only the mass density is considered. These results demonstrate that including the full dynamical structure of the DM spike is essential for accurately modeling the backreaction of DM on black hole spacetimes. 
\end{abstract}
\maketitle

\section{introduction}

Dark matter (DM) constitutes a dominant but invisible component of the Universe, inferred from its gravitational influence on galactic rotation curves, gravitational lensing, and the dynamics of large-scale structures \cite{Bertone:2016nfn}. Despite its overwhelming abundance, the microscopic nature of DM remains elusive \cite{Feng:2010gw,Cirelli:2024ssz,Misiaszek:2023sxe}. In the absence of direct detection, one promising avenue to probe DM properties is through its gravitational effects in strong-field regimes, particularly in the environments surrounding black holes (BHs) \cite{Gondolo:1999ef,Gondolo:1999gy,Barausse:2014tra,Eda:2014kra, Lacroix:2016qpq}.


When a BH grows adiabatically inside a DM halo, the gradual deepening of its gravitational potential compresses the surrounding DM distribution into a steep density enhancement known as a DM spike \cite{Gondolo:1999ef,Gondolo:1999gy}. In the adiabatic approximation, the BH is assumed to grow on a timescale much longer than the orbital period of DM particles. Under such slow evolution, the system remains close to equilibrium, and the orbital actions --- serving as adiabatic invariants --- are conserved. This conservation allows one to relate each particle's initial orbit in the halo to its final orbit after the BH growth, thereby reconstructing the modified density profile. The resulting spike follows a power-law form whose index depends on the slope of the original DM halo. For instance, a Navarro-Frenk-White (NFW) halo \cite{Navarro:1996gj} with $\gamma=1$ yields a spike index $\gamma_{\rm sp}\approx3/2$, producing a density enhancement of several orders of magnitude near the BH. This framework has since been extended to general-relativistic settings, where the spike structure is computed in the Schwarzschild \cite{Sadeghian:2013laa,Sadeghian:2013bga} or Kerr spacetime \cite{Ferrer:2017xwm}. Relativistic corrections become important close to the event horizon and can alter both the inner slope and the maximum density of the spike. The inclusion of strong-field effects is particularly relevant for modeling potential observational signatures, such as gamma-ray emission from DM annihilation or gravitational waves \cite{Yue:2017iwc,Hannuksela:2019vip,Yue:2019ozq,Speeney:2022ryg,Coogan:2021uqv,Becker:2021ivq,Cardoso:2022whc,Cole:2022yzw,Chan:2022gqd,Dai:2023cft,Rahman:2023sof,Zhang:2024ugv,Montalvo:2024iwq,Zhang:2024hrq,Tan:2024hzw,Zhao:2024bpp,Shadykul:2024ehz,Cheng:2024mgl,Gliorio:2025cbh,Dosopoulou:2025jth,Alloqulov:2025ucf,Alnasheet:2025mtr,Ashoorioon:2025ezk,Bhattacharya:2025lvn}.

In most treatments, however, the DM spike has been modeled under simplified assumptions about its energy-momentum content. Many analyses approximate the system as an ideal or anisotropic fluid characterized mainly by its rest-mass density, while neglecting kinetic and anisotropic pressure terms. In the simplest cases, the radial pressure is set to zero, corresponding to circular orbits, whereas more general approaches allow for a tangential component to represent the ensemble-averaged orbital motion of DM particles \cite{Xu:2018wow,Cardoso:2021wlq,Konoplya:2022hbl,Daghigh:2022pcr,Shen:2023erj,Acharyya:2023rnq,Zhang:2024hjr,Shen:2024qbb,Maeda:2024tsg,Chakraborty:2024gcr, Fernandes:2025osu}. Although these approximations greatly simplify the Einstein equations and are sufficient for estimating the density profile, they omit part of the stress-energy associated with the distribution of bound geodesic orbits. Such missing contributions are responsible for the kinetic and anisotropic pressures that also gravitate and hence can influence the spacetime geometry. Including these effects allows one to assess the backreaction of the DM spike on the metric --- a key step toward a more self-consistent description of BHs embedded in dense DM environments \cite{Zhang:2021bdr,Zhang:2022roh,Figueiredo:2023gas,Zhao:2023tyo,Zhao:2023itk,Daghigh:2023ixh,Speeney:2024mas,Liu:2024xcd,Pezzella:2024tkf,Toshmatov:2025rln,Liu:2025otw,Rahman:2025mip,Li:2025eln,Das:2025vja}. 

To incorporate these dynamical effects more consistently, we adopt the Einstein cluster model \cite{Einstein:1939ms,Comer:1993a}, which describes a self-gravitating system of collisionless particles with an anisotropic velocity distribution. In our framework, the DM spike is assumed to form under adiabatic evolution. This allows the spike to be constructed self-consistently from an initial DM halo through conservation of orbital actions.
For the host halo, we adopt a Hernquist profile \cite{Hernquist:1990be}, which provides a realistic finite-mass model with an inner slope similar to that of the NFW profile while offering analytic simplicity. The corresponding spike distribution function (DF) is then used to compute the full energy-momentum tensor of the DM spike from the particle distribution function, explicitly accounting for the kinetic contribution to the energy density and the nonvanishing pressure components arising from noncircular orbits. This approach bridges the gap between oversimplified fluid models and the fully self-consistent Einstein-Vlasov description, while remaining tractable for numerical implementation. Using the derived energy-momentum tensor as a fixed source, we numerically solve the Einstein field equations for a static, spherically symmetric spacetime containing a BH surrounded by a DM spike. By using the parameters for the Milky way fitted in \cite{Eadie:2015}, we find that the kinetic term enhances the total energy density near the spike by roughly 50\% compared with the rest-mass contribution, while the radial pressure, though subdominant, introduces mild anisotropy in the stress components. We find the derived tensor satisfies all the standard energy conditions. Furthermore, the resulting spacetime  exhibits a slight but non-negligible deviation from the Schwarzschild solution --- roughly a factor of two larger than that found in previous treatments considering only the mass density.  Although our model is semiconsistent, whose energy-momentum tensor and metric are solved iteratively rather than solved simultaneously, this approximation is physically well-justified. In realistic systems, the DM spike around a supermassive BH is expected to be extremely dilute, leading to a weak gravitational backreaction. We begin with an analysis within a perturbative framework and find that the iterative treatment captures the leading-order effects of the DM spike. This treatment therefore represents a realistic step toward a fully self-consistent relativistic description of BHs embedded in DM halos.

The remainder of this paper is organized as follows.
Section \ref{section DM spike} reviews the theoretical framework used to describe DM spikes formed under adiabatic evolution and derives the full energy-momentum tensor of the spike from the DF constructed for a Hernquist-type halo. Section \ref{section BH in the DM spike} derives the Einstein field equation for solving a static, spherically symmetric solution and imposes the boundary conditions. The results are presented in Sec. \ref{sec:results}. Section \ref{discuss} summarizes our conclusions and outlines possible extensions. The geometric units $G=c=1$ are maintained throughout this paper except when specifically mentioned.


\section{Theoretical framework for DM spikes}
\label{section DM spike}

\subsection{Distribution function}
Within the framework of classical mechanics, the phase-space DF $f^{(3)}$ for a self-gravitating system can be defined as
\begin{equation}
    \int  f^{(3)}(\vec{r},\vec{p}) \ d^{3}r \  d^{3}p = 1, 
    \label{eq:DF}
\end{equation}
where the integral extends over all phase space $(\vec{r},\vec{p})$. In other words, $f^{(3)}(\vec{r},\vec{p})$ describes the probability density for DM particles at specified phase-space coordinates.  The superscript $(3)$ of the DF represents that the particle orbit are evaluated in the nonrelativistic framework. We denote the DM particle mass, total particle number and total mass of this system as $\mu$, $N$, and $M$, respectively. These three parameters are related to each other by the formula
\begin{equation}
    M=\mu N .
    \label{eq:M mu N}
\end{equation}
The mass density profile $\rho_{M}(r)$ of this system is then expressed as the integral over the momentum space
\begin{equation}
    \rho_{M}(\vec{r}) \equiv \int \left[M f^{(3)}(\vec{r},\vec{p})\right] \ d^{3}p ,
    \label{eq: classical mass density}
\end{equation}
since $M f^{(3)}(\vec{r},\vec{p})$ represents the mass density in the phase space. The mean value of any quantity $Q$ can be imposed by
\begin{equation}
    \langle Q \rangle \equiv \frac{1}{n(\vec{r})} \int Q \left[N f^{(3)}(\vec{r},\vec{p})\right] \ d^{3}p, 
    \label{eq:mean value}
\end{equation}
where $n(\vec{r}) \equiv \int \left[N f^{(3)}(\vec{r},\vec{p})\right] \ d^{3}p$ denotes the number density in the coordinate space. For the relativistic cases, the necessary changes consist of modifying the volume element $d^{3}p$ to $\sqrt{-g} \  d^{4}p$ and performing the integral in the mass shell ($p^{\mu}p_{\mu}=-\mu^{2}$ for the mass particle) in phase space. 

Note that DFs in the literature often represent not the probability density in phase space, but rather the number density or mass density. These distinctions merely correspond to multiple the  probability density $f$ by $N$ or $M$, respectively. In this paper, all DFs in the following of this work are referred to  probability density at specified phase-space coordinates. While citing a DF from other works, we will explicitly indicate their physical meaning (probability density, particle number density, or mass density) to avoid confusion.

For a galactic system with a spherical gravitational potential $\Phi(r)$, the Hamiltonian of a DM particle per unit mass can be expressed as $H \equiv \frac{1}{2} v^2 + \Phi$, and we define the relative potential $\Psi$ and the relative energy $\mathcal{E}$ for a particle per unit mass by
\begin{equation}
    \Psi \equiv -\Phi, \ \ \mathcal{E} \equiv -H = \Psi-\frac{1}{2} v^2,
    \label{eq:relative potential and energy}
\end{equation}
where $v$ is the velocity of the particle.
The relative potential is fixed to vanish at infinity and non-negative for any $r$, while the relative energy is hence equal to the binding energy of a particle. As we only consider bound particles for this system, $\mathcal{E}$ ranges from $0$ to $\Psi(r)$ for a given position.
It is worth noting that the DF $f^{(3)}$ is ergodic for this spherical isotropic system (\ref{eq:relative potential and energy}), i.e., $f^{(3)}$ can be expressed solely as a function $f^{(3)}(\mathcal{E})$ of energy $\mathcal{E}$ \cite{Binney:1987}. 

From (\ref{eq:relative potential and energy}) and the definition $\vec{p}\equiv \mu \vec{v}$, the Eq. (\ref{eq: classical mass density}), for a spherical and isotropic system, can be written as
\begin{eqnarray}
    \rho_{M} &=& \int \left[\mu^{3} M f^{(3)}(\mathcal{E})\right] d^{3}v \nonumber \\ 
                      &=& 4\pi \int \left[\mu^{3} M f^{(3)}(\mathcal{E})\right] v^{2} dv\nonumber \\
                      &=& 4 \sqrt{2} \pi \int^{\Psi}_{0} \left[\mu^{3} M f^{(3)}(\mathcal{E})\right] \sqrt{\Psi-\mathcal{E}} \  d\mathcal{E}
    \label{eq:f to rho 2}
\end{eqnarray}
and the derivation gives
\begin{equation}
    \frac{1}{ \sqrt{8} \pi } \frac{d \ \rho_{M}}{d\Psi} = \int^{\Psi}_{0} d\mathcal{E} \frac{\mu^{3} M f^{(3)}(\mathcal{E})}{\sqrt{\Psi-\mathcal{E}}},
    \label{eq:d f to rho}
\end{equation}
which is an Abel integral equation having a solution
\begin{equation}
    \mu^{3} M f^{(3)}(\mathcal{E}) = \frac{1}{ \sqrt{8} \pi^2} \frac{d}{d\mathcal{E}} \int^{\mathcal{E}}_{0} \frac{d\Psi}{\sqrt{\mathcal{E}-\Psi}} \frac{d\rho_{M}}{d\Psi}.
    \label{eq:rho to f}
\end{equation}
The above Eq. (\ref{eq:rho to f}) imposes the standard formula for transforming a observed (or numerically simulated) mass density to the DF for the spherical and isotropic system. This method is known as Eddington inversion \cite{Binney:1987}.

In this paper, for the DM halo we employ the Hernquist profile \cite{Hernquist:1990be}
\begin{equation}
        \rho_{M}^{(H)}(r) = \frac{M}{2\pi} \frac{a}{r} \frac{1}{(r+a)^{3}}, 
    \label{eq:Hernquist profile}
\end{equation}
where $a$ is a length scale denoting the radius containing a mass of $M/4$. The parameters of the Hernquist model for the Milky Way are fitted in \cite{Eadie:2015}, which gives $M=1.55 \times 10^{12} M_{\odot}$ and $a=16.9 \,\textrm{kpc}$. We adopt the mass of the BH at the center of the Milky Way as $M_{BH}=4\times 10^{6} M_{\odot}$. In the geometric units, the ratio of the above quantities is 
\begin{equation}
    M = 387500 \ M_{BH}, \ \ a = 227775 \ M.
    \label{eq:Milky Way}
\end{equation}
The corresponding relative potential and the mass distribution of the Hernquist profile is given by
\begin{equation}
    \Psi^{(H)}(r) = \frac{M}{r+a}, \ \ M^{(H)}(r) = \frac{M r^2}{(r+a)^2},
    \label{eq:Hernquist potential and mass distribution}
\end{equation}
with $M(a)=M/4$. 
For the Hernquist profile (\ref{eq:Hernquist potential and mass distribution}), the DF $f^{(H)}(\mathcal{E})$ is obtained as
\begin{eqnarray}
    \mu^{3} M f^{(H)}(\mathcal{E}) &=& \frac{\sqrt{\mathcal{E}} (2 a \mathcal{E}-M) \left(-8 a^2 \mathcal{E}^2+8 a M \mathcal{E}+3 M^2\right)}{8 \sqrt{2} \pi ^3 a M^3 (a \mathcal{E}-M)^2} \nonumber \\
    && +\frac{3 M^4  \tan ^{-1}\left(\sqrt{\frac{a\mathcal{E}}{M-a \mathcal{E}}}\right)}{8 \sqrt{2} \pi ^3 a M^3 (M-a \mathcal{E})^{5/2}}.
    \label{eq:Hernquist f}
\end{eqnarray}
Here we regain the result presented by Hernquist \cite{Hernquist:1990be}.

We adopt the Hernquist profile primarily because it possesses a known analytical DF (\ref{eq:Hernquist f}). While other profiles could, in principle, be used, they pose no fundamental difficulty to our calculations. The pair of equations (\ref{eq:f to rho 2}) and (\ref{eq:rho to f}) provides a crucial bridge between the microscopic dynamics, described by the distribution function $f(\mathcal{E})$, and the macroscopic density profile $\rho(r)$ observed in simulations or astronomical observations.

\subsection{Adiabatic growth of a BH}
In the previous subsection, the motion of bound DM particles is completely determined by their self-gravity with no central BH present. The actions of the particles can be defined as
\begin{eqnarray}
    I^{(0)}_{r} &\equiv \oint v_{r} dr&  = \oint \sqrt{2\Psi-2\mathcal{E}-L^{2}/r^{2}},  \nonumber \\
    I^{(0)}_{\theta} &\equiv \oint v_{\theta} d\theta & = 2\pi (L-L_{z}),  \nonumber \\
    I^{(0)}_{\varphi} &\equiv \oint v_{\varphi} d\varphi & = 2\pi L_{z},
    \label{eq:Hernquist adiabatic invariants}
\end{eqnarray}
where $v_{i}$ ($i=r,\theta,\varphi$) represent the velocity's components. However, the presence of a supermassive BH at the center of most galaxies must be accounted for.  A effective approach is adiabatically growing a BH within an initial halo  that lacks one. This process, though it alters the central gravitational potential, is assumed to occur over timescales much longer than the typical orbital period of DM particles. Under this adiabatic approximation, the actions of the particles Eq. (\ref{eq:Hernquist adiabatic invariants}) remain constant as the potential changes. These conserved quantities are known as adiabatic invariants. 

We now proceed to construct the adiabatic invariants for DM particles orbiting a central BH within the halo. However, this task is not straightforward. The standard formulation of adiabatic invariants is derived within Newtonian mechanics, which presents a conceptual challenge since a BH is an inherently relativistic object. Consequently, the orbits of bound particles in its vicinity must be reexamined. Following \cite{Sadeghian:2013laa}, we assume that the geodesics of DM particles are entirely governed by the Schwarzschild metric of the BH. The validity of this assumption will be assessed in the following subsection.

The line element of the Schwarzschild BH is given by
\begin{equation}
    ds^2 = -\left(1-\frac{2M_{BH}}{r}\right) dt^2 +\frac{dr^2} {1-\frac{2M_{BH}}{r}}  + r^2 d\Omega^2,
    \label{eq:ansatz Schwarzschild}
\end{equation}
where $d\Omega^2=d\theta^2+\sin^2\theta d\varphi^2$.
The complete integrability of timelike geodesics in the spacetime (\ref{eq:ansatz Schwarzschild}) ensures that they can be described completely by four orbital conservation quantities; the mass of the DM particles $\mu$, the energy per unit mass $E$, the angular momentum per unit mass $L$ and its $z$-components $L_{z}$,
\begin{eqnarray}
    \mu^2 &=& -p^{\nu}p_{\nu}, \label{eq:particle mass} \\
    E &=& -u_{t}, \label{eq:particle energy} \\
    L_{z} &=& u_{\varphi}, \label{eq:particle z angular momentum} \\
    L^{2} &=& u_{\theta}^2  + \frac{u_{\varphi}^{2}}{\sin^2 \theta}, \label{eq:particle total angular momentum} 
\end{eqnarray}
where $p_{\nu} \equiv \mu u_{\mu}$ represents the 4-momentum of the DM particles.

From (\ref{eq:particle mass})-(\ref{eq:particle total angular momentum}), one can express the radial effective potential of the DM particles as
\begin{eqnarray}
    V_{eff} &\equiv& -u_{r}^{2}\left( 1-\frac{2M_{BH}}{r}\right)^2 \nonumber \\
    &=&  -E^2 + \left( \frac{ r-2 M_{BH}}{r}\right) \left( 1+\frac{L^{2}}{r^2} \right) .
    \label{eq:effective potential}
\end{eqnarray}
Intuitively, for a given energy and angular momentum, the radial motion of a DM particle is permitted only where the effective potential Eq. (\ref{eq:effective potential}) is nonpositive. Such particles can either plunge into the horizon, escape to infinity, or be confined between an inner and an outer turning point. Equation (\ref{eq:effective potential}) indicates that the energy of a bound particle must be less than unity. Furthermore, since bound particles do not cross the horizon, their energy also has a lower bound. The characteristic trajectory of a bound particle is elliptical orbits, oscillating periodically between its pericenter and apocenter. A circular orbit represents a special case of this, where the pericenter and apocenter coincide.

Similar to  (\ref{eq:Hernquist adiabatic invariants}), the adiabatic invariants of the timelike geodesic in the Schwarzschild spacetime can also be constructed by the 4-velocity components,
\begin{eqnarray}
    I^{(1)}_{r} &\equiv \oint u_{r} dr&= \oint dr \left( 1-\frac{2M_{BH}}{r}\right)^{-1} \sqrt{-V_{eff}} ,  \nonumber \\
    I^{(1)}_{\theta} &\equiv \oint u_{\theta} d\theta&  = 2\pi (L-L_{z}),  \nonumber \\
    I^{(1)}_{\varphi} &\equiv \oint u_{\varphi} d\varphi&  = 2\pi L_{z},  
    \label{eq:Schwarzschild adiabatic invariants}
\end{eqnarray}
where the second equality in each row uses (\ref{eq:effective potential}), (\ref{eq:particle total angular momentum}) and (\ref{eq:particle z angular momentum}), respectively. The adiabatic growth of a BH in the DM halo requires the equality between the initial and final adiabatic invariants throughout the slow evolution of the potential
\begin{equation}
    I^{(0)}_{r}=I^{(1)}_{r}, \quad I^{(0)}_{\theta}=I^{(1)}_{\theta}, \quad I^{(0)}_{\varphi}=I^{(1)}_{\varphi}.
    \label{eq:equality of adiabatic invariants}
\end{equation}
By comparing (\ref{eq:Hernquist adiabatic invariants}) and (\ref{eq:Schwarzschild adiabatic invariants}), a simple conclusion is that the angular momentum $L$ and its $z$-component of a DM particle remains unchanged during the adiabatic growth, while its energy must vary. The initial energy $\mathcal{E}$ and final energy $E$ during the adiabatic growth can be related through
\begin{equation}\
    \begin{array}{ll}
    \oint dr\sqrt{2\Psi-2\mathcal{E}-\frac{L^{2}}{r^{2}}} = &   \\
    \oint dr \left( 1-\frac{2M_{BH}}{r}\right)^{-1} \sqrt{E^2 - \left( 1-\frac{2M_{BH}}{r}\right) \left( 1+\frac{L^{2}}{r^2} \right)}. & 
    \end{array}
    \label{eq:energy in adiabatic growth}
\end{equation}
This relation generates an implicit function $\mathcal{E}=\mathcal{E}(E,L^{2})$ for the initial particle energy which can be evaluated numerically.

\subsection{DM spike around a BH}

Research \cite{Gondolo:1999ef,Gondolo:1999gy,Sadeghian:2013bga,Sadeghian:2013laa} indicates that a Schwarzschild BH leads to the formation of a high-amplitude DM density spike, which can be several orders of magnitude denser than the initial Hernquist profile. Additionally, a Kerr BH's spin is known to further increase the DM density near the horizon \cite{Ferrer:2017xwm}. Building upon this foundation, we begin by reviewing the derivation of the DM density profile influenced by a Schwarzschild BH. The primary new contribution of this subsection is the first presentation of the energy-momentum tensor for this DM profile following adiabatic BH growth. This step is essential because the energy-momentum tensor serves as the source term that couples matter in the Einstein field equation.

\begin{figure}[htbp]
	\centering
	\includegraphics[width = 0.450\textwidth]{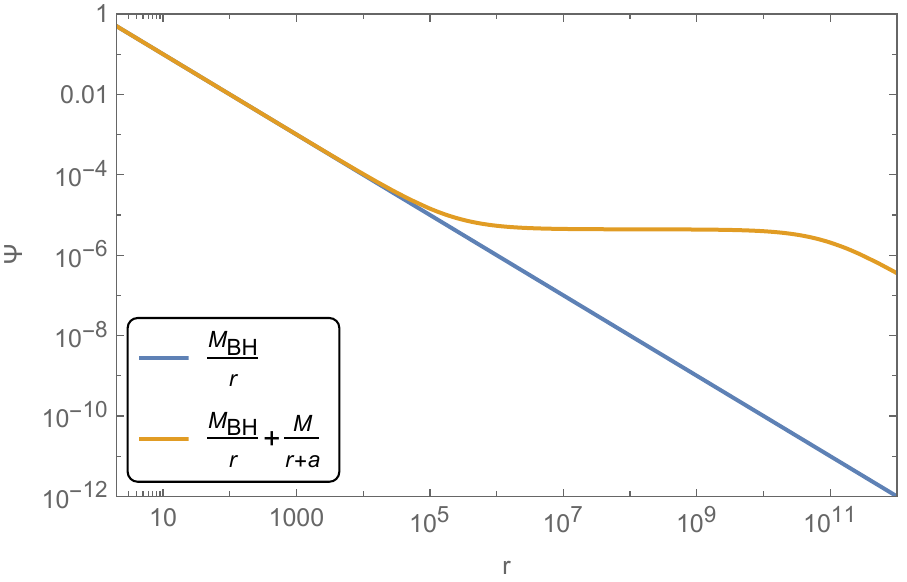}
	\caption{The relative potential for a BH and an isolated point mass, respectively. This plot shows that these two potentials are similar before $r=10^{5}$, where we set $M_{BH}=1$.}
	\label{fig:Mvsr}
\end{figure}

Given our focus on the backreaction of the DM spike on the BH, a simplified treatment of the spike profile is warranted. We adopt the standard approach from \cite{Gondolo:1999ef,Gondolo:1999gy,Sadeghian:2013bga,Sadeghian:2013laa}, assuming the geodesics of DM particles are governed solely by the Schwarzschild metric of the central BH. Our calculation is confined to the radial interval $r_{H} \leq r \leq 10^5$, where $r_{H}$ is the event horizon radius and we set $M_{BH}=1$. This restriction is justified by
Fig. \ref{fig:Mvsr}, which shows that within this range, the total potential $\Psi$ is dominated by the BH and is indistinguishable from that of an isolated point mass. Furthermore, the enclosed mass of the Hernquist profile at $r = 10^5$  is negligible compared to the BH mass, validating the point-mass approximation for the BH's potential at these distances.

For the relativistic galactic dynamics, the mass current density $j^{\mu}$ of identical particles for a DM distribution is defined by $j_{\mu} \equiv \rho_{M}(x) \langle u_{\mu} \rangle$. From (\ref{eq:mean value}), this can be expressed by
\begin{equation}
    j_{\mu}(x)  = \int u_{\mu} \ \left[M f^{(4)}(x,p)\right] \ \sqrt{-g^{(S)}} \ d^{4}p, 
    \label{eq:4 flux on momentum space} 
\end{equation}
where the integral is performed in the mass shell $p^{\mu}p_{\mu} = -\mu^{2}$. The function $M f^{(4)}(x,p)$ represents the mass density of DM particles at specified phase-space coordinates, and $\sqrt{-g^{(S)}} = r^2 \sin \theta $ for the Schwarzschild BH. The superscript $(4)$ means that $ f^{(4)}(x,p)$ is a relativistic DF different from its nonrelativistic counterpart $f^{(3)}(\vec{r},\vec{p})$. 

The energy-momentum tensor $T^{\mu}{}_{\nu}$ of a galactic system has been statistically established by Einstein, known as the Einstein cluster \cite{Einstein:1939ms}. Following this construction, we express the energy-momentum tensor as
\begin{equation}
    T_{\mu}{}^{\nu}(x) \equiv    \frac{n(x)}{\mu} \langle p_{\mu}p^{\nu} \rangle, 
    \label{eq:Einstein cluster}
\end{equation}
where $n(x)$ represents the particle number density at specified coordinates. The marker $\langle p_{\mu}p^{\nu} \rangle$ denotes the mean value of $p_{\mu}p^{\nu}$ for DM particles at spacetime coordinate $x$, which can be expressed as $ \frac{1}{n(x)}  \int p_{\mu}p^{\nu} \ \left[N f^{(4)}(x,p)\right] \ \sqrt{-g^{(S)}} \ d^{4}p$ via the DF. The energy-momentum tensor is then given by
\begin{equation}
    T_{\mu}{}^{\nu}(x) =   \int u_{\mu}u^{\nu} \  \left[M f^{(4)}(x,p)\right] \  \sqrt{-g^{(S)}} \ d^{4}p,
    \label{eq:energy momentum tensor on momentum space}
\end{equation}
using (\ref{eq:M mu N}). The expressions (\ref{eq:4 flux on momentum space}) and (\ref{eq:energy momentum tensor on momentum space}) align with the form of (16) and (17) in relativistic stellar dynamics \cite{Fackerell:1968}, where the DF ``f'' represents the particle number density in phase space and corresponds  to $N f^{(4)}(x,p)$ in this paper. 

Following \cite{Fackerell:1968}, the DF $f^{(4)}$ further takes the form, 
\begin{equation}
    f^{(4)}(x,p) = f(E,L^{2}) \delta(\mu'-\mu),
    \label{eq:relativistic distribution}
\end{equation}
where the Dirac delta function is generated from DM particles with identical mass, and the phase space distribution $f(E,L^{2})$ has been proven to preserve the form for the adiabatic growth of a BH within the DM halo 
\begin{equation}
    f(E,L^{2}) = f^{(H)} \left[ \mathcal{E}(E,L^{2}) \right],
    \label{eq:adiabatic DF}
\end{equation}
along with the transformation (\ref{eq:energy in adiabatic growth}) \cite{Sadeghian:2013laa,Sadeghian:2013bga}. Given that $f(E,L^{2})$ depends solely on the conserved quantities, it is convenient to transform the integral from the momentum space into the conserved quantity space. The volume element of the momentum space in integrations (\ref{eq:4 flux on momentum space}) and (\ref{eq:energy momentum tensor on momentum space}) can then be converted into
\begin{eqnarray}
    \sqrt{-g^{S}} d^{4} p &=& \sqrt{-g^{S}} \left| \frac{\partial(p^{t},p^{r},p^{\theta},p^{\varphi})}{\partial(\mu',E,L^{2},L_{z})} \right| d\mu'dE dL^{2} dL_{z} \nonumber \\
            &=& \frac{\mu'^{3}}{2r^{2} |u^{r}| |u_{\theta}| \sin\theta} d\mu'dE dL^{2} dL_{z}, \label{eq:volume element}
\end{eqnarray}
where 
\begin{equation}
    u^{r} = \pm \sqrt{-V_{eff}} \ , \ u_{\theta} = \pm \sqrt{L^{2}-L_{z}^{2}\sin^{-2}\theta}, 
    \label{eq:ur and utheta}
\end{equation}
read from (\ref{eq:particle total angular momentum}) and (\ref{eq:effective potential}). 

\begin{figure}[htbp]
	\centering
	\includegraphics[width = 0.450\textwidth]{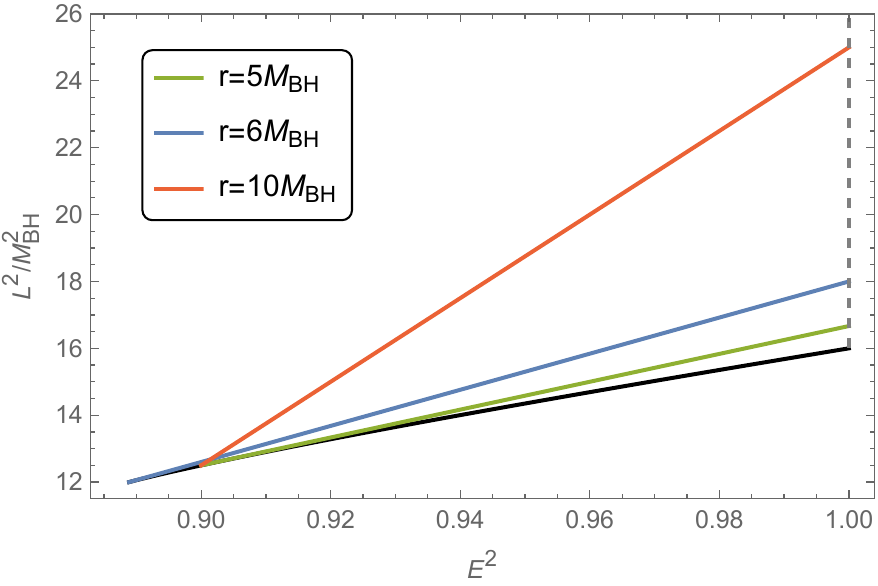}
	\caption{This plot shows the integration region $(E,L)$  of integrals (\ref{eq:4 flux on considered quantities}) and (\ref{eq:energy momentum tensor on considered quantities}) for each fixed $r$. The colored lines represent the maximal $L^2$ for a given $r$ and $E$.}
	\label{fig:region E2L2}
\end{figure}

Combining (\ref{eq:relativistic distribution}), (\ref{eq:adiabatic DF}) and (\ref{eq:volume element}), the mass current density (\ref{eq:4 flux on momentum space}) and the energy-momentum tensor (\ref{eq:energy momentum tensor on momentum space}) can be written as
\begin{eqnarray}
    j_{\mu}      &=&  \int u_{\mu} \left( \mu^{3}M f^{(H)}\left[ \mathcal{E}(E,L^{2}) \right]\right) \nonumber \\
    && \times \frac{dE \ dL^{2} dL_{z}}{2r^{2} |u^{r}| |u_{\theta}| \sin\theta} , \label{eq:4 flux on considered quantities} \\
    T_{\mu}{}^{\nu} &=& \int u_{\mu}u^{\nu} \left( \mu^{3}M f^{(H)}\left[ \mathcal{E}(E,L^{2}) \right]\right) \nonumber \\
    && \times \frac{dE \ dL^{2} dL_{z}}{2r^{2} |u^{r}| |u_{\theta}| \sin\theta} . \label{eq:energy momentum tensor on considered quantities}
\end{eqnarray}
It is remarkable that, although we begin from the $(x,p)$ phase space, the resulting expressions for the mass current density Eq. (\ref{eq:4 flux on considered quantities}) and the energy-momentum tensor Eq. (\ref{eq:energy momentum tensor on considered quantities}) are independent of the DM particle mass $\mu$, except for the term $\mu^{3}M f^{(3)}\left[ \mathcal{E}(E,L^{2}) \right]$. The latter, however, corresponds exactly to the DF Eq. (\ref{eq:Hernquist f}) of the Hernquist profile, which depends only on the macroscopic parameters ($M$, $a$, $\mathcal{E}$). Therefore, the calculation does not require any knowledge of the specific DM particle; it relies solely on macroscopic parameters --- obtained from observations or numerical simulations --- together with the orbital properties of a unit-mass timelike particle near the BH. This approach circumvents the need for a specific DM particle model and offers a universal description of the DM spike around a BH.

The integration ranges in Eqs.~\eqref{eq:4 flux on considered quantities} and~\eqref{eq:energy momentum tensor on considered quantities} are defined by the set of all bound timelike geodesics in the Schwarzschild background given by Eq.~\eqref{eq:ansatz Schwarzschild}. As specified by Eqs.~\eqref{eq:particle z angular momentum} and~\eqref{eq:particle total angular momentum}, the integral over \( L_z \) for fixed \( L \) and \( \theta \) is taken over the symmetric interval \( (-L \sin\theta, L \sin\theta) \).
The critical (minimum) and maximum values of the total angular momentum \( L^2 \) are given, respectively, by
\begin{eqnarray}
    L^2_{\mathrm{cri}} &=& \frac{32 M_{BH}^{2}}{-27 E^4 + 36 E^2 + E (9 E^2 - 8)^{3/2} - 8}, \label{eq:critical L2} \\
    L^2_{\mathrm{max}} &=& \frac{E^2 r^3}{r - 2 M_{BH} } - r^2. \label{eq:maximal L2}
\end{eqnarray}
The critical angular momentum \( L^2_{\mathrm{cri}} \) is represented by the black curve in the \( (E^2, L^2) \) plane of Fig.~\ref{fig:region E2L2}, while \( L^2_{\mathrm{max}} \) is indicated by colored curves for specific radii; green for \( r = 5 M_{BH}\), blue for \( r = 6 M_{BH}\) (the innermost stable circular orbit), and red for \( r = 10 M_{BH}\). The gray dashed line in the same figure marks the bound condition \( E^2 \leq 1 \), indicating that the particle lacks sufficient energy to escape to infinity.
Thus, for a given radius \( r \), the integration domain in the \( (E^2, L^2) \)-plane is the region enclosed by the corresponding maximal \( L^2 \) curve (green, blue, or red), the critical \( L^2 \) curve (black), and the horizontal line \( E^2 = 1 \) (gray dashed). The intersection point of the critical \( L^2 \) curve and the maximal \( L^2 \) curve defines the minimum energy \( E \) within the integration domain for a given \( r \), which can be expressed as
\begin{equation}
	E_{min} =
	\left\{
	\begin{array}{ll}
		 (1-\frac{2M_{BH}}{r})\cdot(1-\frac{3M_{BH}}{r})^{-\frac{1}{2}} , & 4 \leq \frac{r}{M_{BH}} \leq 6, \\
		 & \\
		 (1+\frac{2M_{BH}}{r})\cdot(1+\frac{6M_{BH}}{r})^{-\frac{1}{2}} , &  \frac{r}{M_{BH}}\geq 6 .
	\end{array}
	\right.
	\label{eq:boundary}
\end{equation}
Among the remaining variables in the volume element (\ref{eq:volume element}), the variable $\mu'$ is constrained (or fixed) by the delta function in (\ref{eq:relativistic distribution}) and is thus integrated out. The minimum $r/M_{BH}=4$ represents the marginally bound orbit, where specifies the minimal orbital radius limit of DM particles. We refer readers to reference \cite{Sadeghian:2013bga,Sadeghian:2013laa} for further details of the derivation.

We now examine each component of \( j_{\mu} \). As shown in Eq.~(\ref{eq:ur and utheta}), the \( r \)- and \( \theta \)-components of the 4-velocity \( u_{\mu} \) each carry a \( \pm \) sign for a given set of \( (E, L^2, L_z) \). Since the integration runs over both positive and negative values of \( u_r \) and \( u_\theta \), these contributions cancel, and thus \( j_r \) and \( j_\theta \) in Eq.~(\ref{eq:4 flux on considered quantities}) vanish. For the \( \varphi \)-component, we note that \( u_{\varphi} = L_z \), and the integration over \( L_z \) spans the symmetric interval \( (-L \sin\theta, L \sin\theta) \). Since the integrand in \( j_\varphi \) is an odd function of \( L_z \), integration over this symmetric range also yields zero. 
Therefore, the only nonvanishing component of Eq.~(\ref{eq:4 flux on considered quantities}) is 
\begin{widetext}
\begin{eqnarray}
    j_{t} &=& 4 \int (-E) \times \frac{  \mu^{3}M f^{(H)}\left[ \mathcal{E}(E,L^{2}) \right]}{2r^{2} \sqrt{E^2 - \left( \frac{ r-2 M_{BH}}{r}\right) \left( 1+\frac{L^{2}}{r^2} \right)}  \sqrt{L^{2}\sin^{2}\theta-L_{z}^{2}} } dE dL^{2} dL_{z} \nonumber \\
          &=& -4\pi \int^{1}_{E_{min}} E \ dE \int^{L_{max}}_{L_{cri}}L dL
          \frac{ \mu^{3}M f^{(H)}\left[ \mathcal{E}(E,L^{2}) \right]}{r^2 \sqrt{E^2 - \left( \frac{ r-2 M_{BH}}{r}\right) \left( 1+\frac{L^{2}}{r^2} \right)}},
    \label{eq:jt}
\end{eqnarray}
\end{widetext}
where the extra factor of $4$ in front of the integral in the first line is attributed to the inclusion of both positive and negative values of $u_{r}$ and $u_{\theta}$. This extra factor $4$ is also included in the following derivation of the energy-momentum tensor. Here we have used (\ref{eq:particle energy}) and (\ref{eq:Hernquist f}).

The analysis of the energy-momentum tensor follows the reasoning previously applied to \( j_\mu \). Components of \( T_{\mu}{}^{\nu} \) containing a single subscript or superscript \( r \) or \( \theta \) vanish upon integration, since the integrand takes both positive and negative values of \( u_r \) and \( u_\theta \). Furthermore, the integral for \( T_{r}{}^{\theta} \) (or \( T_{\theta}{}^{r} \)) reduces to an expression of the form \( \int \left[ 2u_r u_\theta - 2u_r u_\theta \right] \cdots \), which cancels identically. Similarly, the components \( T_{t}{}^{\varphi} \) and \( T_{\varphi}{}^{t} \) vanish for the same reason as \( j_\varphi \): the integrand is an odd function of \( L_z \), and the integration domain is symmetric about zero. As a result, the only nonvanishing components of \( T_{\mu}{}^{\nu} \) are the diagonal ones, which can be expanded as
\begin{widetext}
\begin{eqnarray}
    T_{t}{}^{t}             &=& -2\pi \int^{1}_{E_{min}} E^{2} dE \int^{L_{max}}_{L_{cri}} dL \frac{ \mu^{3}M f^{(H)}\left[ \mathcal{E}(E,L^{2}) \right]}{(r^{2}-2M_{BH}\ r) \sqrt{E^2 - \left( \frac{ r-2 M_{BH}}{r}\right) \left( 1+\frac{L^{2}}{r^2} \right)}}, \label{eq:Ttt} \\
    T_{r}{}^{r}             &=& 2 \pi \int^{1}_{E_{min}} dE \int^{L_{max}}_{L_{cri}} dL \sqrt{E^2 - \left( \frac{ r-2 M_{BH}}{r}\right) \left( 1+\frac{L^{2}}{r^2} \right)} \ \  \frac{\mu^{3}M f^{(H)}\left[ \mathcal{E}(E,L^{2}) \right]}{r^{2}-2M_{BH}\ r}, \label{eq:Trr} \\
    T_{\theta}{}^{\theta}   &=& \  T_{\varphi}{}^{\varphi} \ = \pi \int^{1}_{E_{min}} dE \int^{L_{max}}_{L_{cri}} L^{2} dL  \frac{\mu^{3}M f^{(H)}\left[ \mathcal{E}(E,L^{2}) \right]}{r^{4}\sqrt{E^2 - \left( \frac{ r-2 M_{BH}}{r}\right) \left( 1+\frac{L^{2}}{r^2} \right)} }. \label{eq:Tthetatheta and Tvarphivarphi} 
\end{eqnarray}
\end{widetext}
For the non-negative DF (\ref{eq:Hernquist f}), the component $T_{t}^{t}$ is nonpositive for $r>r_{H}=2M_{BH}$, while the remaining diagonal components are both non-negative, which indicates that the system exhibit positive energy and positive pressure. 

\section{Spacetime of the BH embedded in the DM spike}

\label{section BH in the DM spike}

In this section, we investigate the backreaction of a DM spike on the BH. We begin with a perturbative analysis, in which the DM spike is treated as a small perturbation, and review the Einstein field equations for a standard static, spherically symmetric ansatz. We then discuss the boundary conditions for solving these equations and briefly describe the numerical method.

\begin{figure}[htbp]
	\centering
	\includegraphics[width = 0.450\textwidth]{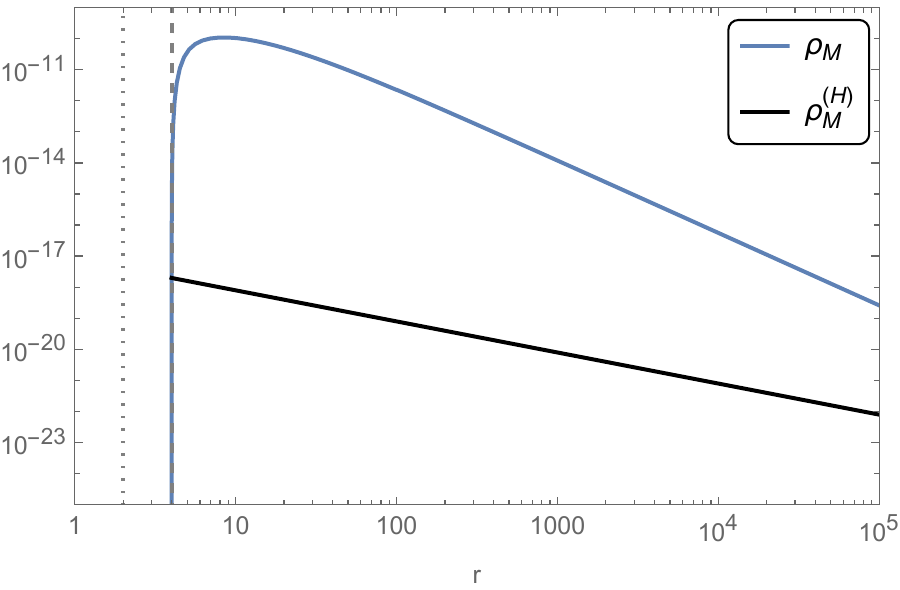}
    \includegraphics[width = 0.450\textwidth]{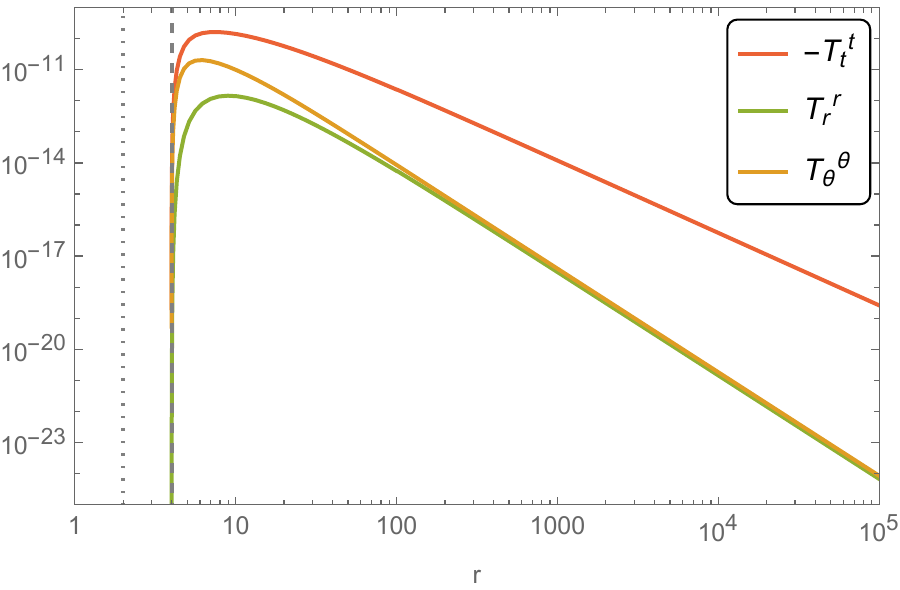}
	\caption{The mass density (upper) and energy-momentum tensor (bottom) for the DM profile around a Schwarzschild BH. The black line in the upper panel shows the initial mass density for the Hernquist profile. After the adiabatic growth of a BH, $\rho_{M}$ forms a ``spike'' near the BH. Components of $T_{\mu}^{\nu}$ exhibit a similar spike, although their peak values differ by orders of magnitude.}
	\label{fig:rho T vs r}
\end{figure}

\subsection{Perturbative framework of Einstein equation}

Using Galactic parameters (\ref{eq:Milky Way}) for the Hernquist profile, the characteristic length of the DM halo is extremely large. Although the total mass of DM halo exceeds the BH mass by five orders of magnitude, the halo mass enclosed within a finite cutoff radius is much smaller than that of the central BH. Consequently, the gravitational field of the BH remains dominant within the cutoff radius, as illustrated by Fig. \ref{fig:Mvsr}, and the gravitational effect of DM halo can be treated as a perturbation. We therefore begin our analysis within a well-established perturbative framework of the Einstein field equation
\begin{eqnarray}
    0&=&G_{\mu\nu}[\mathbf{g}] -T_{\mu\nu}[\rho,\mathbf{g}] , \label{eq:Einstein} \\
    0&=&\nabla^{\mu}T_{\mu\nu}[\rho,\mathbf{g}]  \label{eq:conservation}
\end{eqnarray}
where $G_{\mu\nu}$ is the Einstein tensor, and $T_{\mu\nu}$ represents the energy-momentum tensor of the DM halo. The function $\rho$ encodes the matter field (energy density, radial pressure or tangential pressure) of DM halo, and hence can be naturally expanded $\rho = \epsilon \rho^{(1)} + \epsilon^2 \rho^{(2)} + \ldots$ by the characteristic density ratio $\epsilon \ll 1$. The order of magnitude of $\epsilon \sim\rho_0^{\textrm{DM}}/\rho_0^{\textrm{BH}}$ is estimated to be approximately $10^{-7}$, where $\rho_0^{\textrm{DM}}$ is taken to be the maximum value of the energy density of the DM spike and $\rho_0^{\textrm{BH}}$ is defined as $M_{BH}/r_\textrm{spike}^3$. The backreaction on the spacetime leads to the expansion of the metric $g_{\mu\nu} = g^{(0)}_{\mu\nu} + \epsilon g^{(1)}_{\mu\nu} + \epsilon^2 g^{(2)}_{\mu\nu} + \cdots$. Up to the first order of $\epsilon$, the field equations (\ref{eq:Einstein}) and (\ref{eq:conservation}) can be expanded as 
\begin{eqnarray}
    0&=&G^{(0)}_{\mu\nu}[g^{(0)}], \label{eq:Einstein0} \\
    0&=&G^{(1)}_{\mu\nu}[g^{(0)},g^{(1)}]-8\pi \ T^{(1)}_{\mu\nu}[\rho^{(1)},g^{(0)}], \label{eq:Einstein1} \\
    0&=&\nabla^{\mu}T_{\mu\nu}^{(1)}[\rho^{(1)},g^{(0)}]. \label{eq:conservation1}
\end{eqnarray}
Within this framework, one advantage is that all covariant derivatives and d'Alembertian operators are defined with respect to the background metric. The background equation (\ref{eq:Einstein0}) admits the Schwarzschild solution. The first-order $T^{(1)}_{\mu\nu}[\rho^{(1)},g^{(0)}]$ of the energy-momentum tensor is evaluated on the background of the vacuum Schwarzschild solution, and hence corresponding to the energy-momentum tensor of the DM spike induced by the Schwarzschild BH. Since $T^{(1)}_{\mu\nu}$ is already of order $\epsilon$ from $\rho^{(1)}$, any corrections on the particle momenta induced by $g^{(1)}_{\mu\nu}$ only enter at $\mathcal{O}(\epsilon^2)$. As a result, $T^{(1)}_{\mu\nu}[\rho^{(1)},g^{(0)}]$ can be calculated by expressions (\ref{eq:Ttt})-(\ref{eq:Tthetatheta and Tvarphivarphi}) from the Schwarzschild geodesics. On the other hand, $g^{(1)}_{\mu\nu}$ describes the perturbed metric inspired by $\rho^{(1)}$ from the Eq. (\ref{eq:Einstein1}). The first-order (\ref{eq:conservation1}) of conserved equation is automatically satisfied, indicating that the correction of (\ref{eq:conservation}) induced by the metric backreaction appears in the higher orders. 

In this work, we focus on the first-order effects of the DM spike and neglect all higher-order contributions. For the purpose of conciseness, Eqs. (\ref{eq:Einstein0}) and (\ref{eq:Einstein1}) can be combined
\begin{equation}
    \mathcal{E}_{\mu\nu} \equiv G^{(0+1)}_{\mu\nu}[g_{\mu\nu}]-8\pi \ T^{(1)}_{\mu\nu}[\epsilon \rho^{(1)},g^{(0)}] \approx 0,
    \label{eq:final equation}
\end{equation}
where $ G^{(0+1)}_{\mu\nu}[g_{\mu\nu}] \approx G^{(0)}_{\mu\nu}[g^{(0)}] + \epsilon \  G^{(1)}_{\mu\nu}[g^{(0)},g^{(1)}] + \mathcal{O}(\epsilon^2)$ represents the Einstein tensor constructed by the combined metric $g_{\mu\nu} \equiv g^{(0)}_{\mu\nu}+\epsilon g^{(1)}_{\mu\nu}$. It is remarked that $\epsilon$ is absorbed into the energy-momentum tensor, since it merely serves as a bookkeeping parameter characterizing the perturbative amplitude of the matter field. An advantage of this equation is that it avoids an explicit discussion of gauge freedom for the perturbed metric $g^{(1)}_{\mu\nu}$. Instead, the combined metric $g_{\mu\nu}$ can be calculated through a well-established procedure; assuming a static spherical symmetric ansatz, and solving the  field equation $\mathcal{E}_{\mu\nu}$. The drawback of this equation is the requirement of higher numerical accuracy. Accordingly, all numerical computations are carried out using sextuple machine precision in \textit{Mathematica}.

\subsection{Equations}

The general metric ansatz for a static, spherically symmetric BH is given by
\begin{equation}
    ds^{2} = -h(r)dt^{2} + \frac{dr^{2}}{1-\frac{2m(r)}{r}} +r^{2} d\Omega^2,
    \label{eq:ansatz}
\end{equation}
with two unknown functions $h(r)$ and $m(r)$. 
Conventionally, the DM can be modeled as an anisotropic fluid with the energy-momentum tensor in the comoving frame given by
\begin{equation}
    T^{(1)}{}_{\mu}{}^{\nu} = \textrm{Diag}\{-\rho_{E}(r),p_r(r),p_t(r),p_t(r)\},
    \label{eq:EnergyMomentumTensor}
\end{equation}
where $\rho_{E}(r)$ denotes the energy density of fluid, $p_r(r)$ is the radial pressure and $p_t(r)$ is the tangential pressure. For the conciseness, we omit the marker $(1)$ of $T^{(1)}{}_{\mu}{}^{\nu}$ in the remaining of this paper. The equality $T_{\theta}{}^{\theta} = T_{\varphi}{}^{\varphi}  = p_t$ follows from the spherical symmertry of the spacetime, as can be verified from (\ref{eq:Tthetatheta and Tvarphivarphi}).

Employing the ansatz (\ref{eq:ansatz}), the nonvanishing components of (\ref{eq:final equation}) and (\ref{eq:conservation1}) are given by
\begin{eqnarray}
    \mathcal{E}_{t}{^{t}} &=& 8 \pi  \rho_{E} (r)-\frac{2 m'(r)}{r^2} \label{eq:Ett} \\
    \mathcal{E}_{r}{^{r}} &=& \frac{(r-2 m(r)) h'(r)}{r^2 h(r)}-\frac{2 m(r)}{r^3}-8 \pi  p_{r}(r) \label{eq:Err} \\
    \mathcal{E}_{\theta}{^{\theta}} &=& \frac{(r-2 m(r)) h''(r)}{2 r h(r)}-\frac{h'(r) \left(r \left(m'(r)-1\right)+m(r)\right)}{2 r^2 h(r)} \nonumber \\
     && -\frac{(r-2 m(r)) h'(r)^2}{4 r h(r)^2}+\frac{m(r)-r m'(r)}{r^3}-8 \pi  p_t(r) \nonumber \\ 
     && \label{eq:Ethetatheta} \\
    \mathcal{C}^{(1)}_{r}       &=&  \frac{1}{2} p_r(r) \left(\frac{h'(r)}{h(r)}+\frac{4}{r}\right)+\frac{\rho_{E} (r) h'(r)}{2 h(r)}+p_{r}'(r)-\frac{2 p_t(r)}{r},\nonumber \\ 
     &&  \label{eq:Cr}
\end{eqnarray}
with $\mathcal{E}_{\varphi}{^{\varphi}} = \mathcal{E}_{\theta}{^{\theta}}$. There are four equations with five independent variables. However, we can show that (\ref{eq:Ethetatheta}) is equivalent to (\ref{eq:Cr}). From (\ref{eq:Ett}) and (\ref{eq:Err}), one can find that $m'(r)\to 4 \pi  r^2 \rho_{E} (r)$ and $h(r)\to c_1 \exp \left(\int _{r_h}^r\frac{2 \left[4 \pi  p_r\left(\hat{r}\right) \hat{r}^3+m\left(\hat{r}\right)\right]}{\hat{r} \left[\hat{r}-2 m\left(\hat{r}\right)\right]}d\hat{r}\right)$, here $c_1$ represents an integration constant. Employing such substitutions, (\ref{eq:Ethetatheta}) and (\ref{eq:Cr}) become 
\begin{eqnarray}
    \frac{1}{4\pi r}\mathcal{E}_{\theta}{^{\theta}} &=& \mathcal{C}_{r} = \frac{p_{r}(r) \left(-3 m(r)+4 \pi  r^3 p_{r}(r)+2 r\right)}{r (r-2 m(r))} \nonumber \\
    && +\frac{\rho_{E} (r) \left(m(r)+4 \pi  r^3 p_{r}(r)\right)}{r^2-2 r m(r)}+p_{r}'(r)-\frac{2 p_{t}(r)}{r}. \nonumber \\
    && \label{eq:Cr And Ethetatheta}
\end{eqnarray}
Therefore, the number of independent equations is actually three. 

By substituting the expressions (\ref{eq:Ttt}), (\ref{eq:Trr}) and (\ref{eq:Tthetatheta and Tvarphivarphi}) for the energy-momentum tensor and applying the Schwarzschild metric ($h \rightarrow 1 - 2M_{BH}/r, \ m \rightarrow 1$) to the conservation Eq. (\ref{eq:Cr}), we find that equation (\ref{eq:Cr}) is automatically satisfied, regardless of the specific form of the DF. However, under these conditions, the Eqs (\ref{eq:Ett}), (\ref{eq:Err}), and (\ref{eq:Ethetatheta}) are not satisfied by the Schwarzschild solution, which is known to fulfill only the vacuum Einstein equation (\ref{eq:Einstein0}). Expanding the expression (\ref{eq:energy momentum tensor on momentum space}) of the energy-momentum tensor in a general metric shows that the conservation equation still holds. This indicates that the validity of the conservation equation does not rely on the Einstein field equations themselves, provided that the definition (\ref{eq:energy momentum tensor on momentum space}) of $T_{\mu}{}^{\nu}$ employs the same metric functions ($h, m$) appearing in Eq. (\ref{eq:Cr}).

We compute the energy-momentum tensor of the DM spike in a fixed Schwarzschild BH spacetime and investigate its backreaction on the geometry. In this analysis, Eqs. (\ref{eq:Ett}) and (\ref{eq:Err}) are adopted as constraint equations to determine the metric functions ($h, m$).
The choice is motivated by the fact that the conservation equation (\ref{eq:Cr And Ethetatheta}) is inherently satisfied in the Schwarzschild background, where the DM energy-momentum tensor is constructed via a statistical DF that ensures covariant conservation. Once the backreaction of the DM spike is taken into account, however, the spacetime deviates from Schwarzschild, and this statistical construction no longer guarantees exact conservation, hence (\ref{eq:Cr And Ethetatheta}) cannot be simultaneously fulfilled in the perturbed metric.

In realistic astrophysical systems \cite{Eadie:2015}, the DM spike surrounding a supermassive BH is expected to be extremely dilute, resulting in only a weak gravitational backreaction. Therefore, the iterative treatment adopted here adequately captures the leading order effects of the DM spike, as illustrated in previous subsection.
To fully solve the system of equations (\ref{eq:Ett}), (\ref{eq:Err}) and (\ref{eq:Cr And Ethetatheta}), one would need to compute the energy-momentum tensor in a fully numerical BH background. Such a computation lies beyond the scope of the present work. As will be demonstrated in the next section, the resulting BH solution remains close to the Schwarzschild metric but exhibits a slightly stronger deviation compared to the alternative approach presented in \cite{Chakraborty:2024gcr}.

For comparison, an alternative methodology adopted in \cite{Cardoso:2021wlq,Chakraborty:2024gcr} solves the full system of equations (\ref{eq:Ett}), (\ref{eq:Err}) and (\ref{eq:Cr And Ethetatheta}) by taking $\rho_E$ or $\rho_m$ as an input quantity. This input is typically obtained either by multiplying the DM halo profile by a cutoff factor or by evaluating it in a fixed Schwarzschild BH spacetime within the Einstein cluster framework. In this formulation, the DM spike is modeled as an anisotropic fluid with vanishing radial pressure. Consequently, the three independent variables in the system can be consistently matched to the three independent field equations, enabling a self-consistent treatment of the coupled system.

\subsection{Boundary conditions}

We discuss the boundary conditions for solving Eqs. (\ref{eq:Ett}) and (\ref{eq:Err}). We begin with the simple Eq. (\ref{eq:Ett}), which possesses a straightforward solution $m(r) = \int_{r_{H}}^{\infty} 4 \pi r^{2} \rho_{E}(r)$. The function $m(r)$ is hence named the ``mass function'' for the metric, and represents the energy distribution on the spacetime. Recall that $T_\mu{}^\nu$ vanishes at radius $r_{cri}=4M_{BH}$, $m(r)$ is solely contributed by the BH mass $M_{BH}$ at $r<r_{cri}$. The boundary condition of $m(r)$ at the event horizon is given by
\begin{equation}
    m(r_{H}) = m(r_{cri}) = M_{BH}.
    \label{eq:m boundary condition}
\end{equation}
From $g_{rr} = (1-\frac{2m(r)}{r})^{-1}$, the radius for the event horizon is simply $r_{H} = 2M_{BH}$. Substituting (\ref{eq:m boundary condition}) and $p_{r}=0$ into (\ref{eq:Err}), this equation becomes
\begin{equation}
    (r^{2}-2 M_{BH} \ r) h'(r)=2 M_{BH} \  h(r), \ \ r_{H} \leq r < r_{cri},
    \label{eq:g equation at rcri}
\end{equation}
which gives the solution $h(r) = 1-\frac{2M_{BH}}{r}$ in this interval. One can find that the solutions of metric functions are exactly the Schwarzschild BH for $r<r_{cri}$, which is attributed to the truncation of the DM distribution at $r=r_{cri}$. 

Since the contribution of the DM spike begins at radius $r_{cri}$, we numerically solve the first-order differential equations (\ref{eq:Ett}) and (\ref{eq:Err}) in the interval $ r_{cri} \leq r \leq 10^5$. Boundary conditions at $r_{cri}$ can be imposed 
\begin{equation}
    h(r_{cri}) = 1- \frac{2M_{BH}}{r_{cri}} = \frac{1}{2}, \ \ m(r_{cri}) = M_{BH},
    \label{eq:boundary conditions}
\end{equation}
based on the requirement of continuity.

The system of equations (\ref{eq:Ett}) and (\ref{eq:Err}), subject to the boundary conditions (\ref{eq:boundary conditions}), is solved using a spectral method. We expand the metric functions in a series of Chebyshev polynomials. The governing equations and boundary conditions are then discretized at Chebyshev collocation points, transforming the problem into a system of algebraic equations for the series coefficients. Starting from an initial guess (the Schwarzschild solution in this work), this nonlinear system is solved iteratively via the Newton-Raphson method.

\section{results}
\label{sec:results}

\begin{figure}[htbp]
	\centering
	\includegraphics[width = 0.450\textwidth]{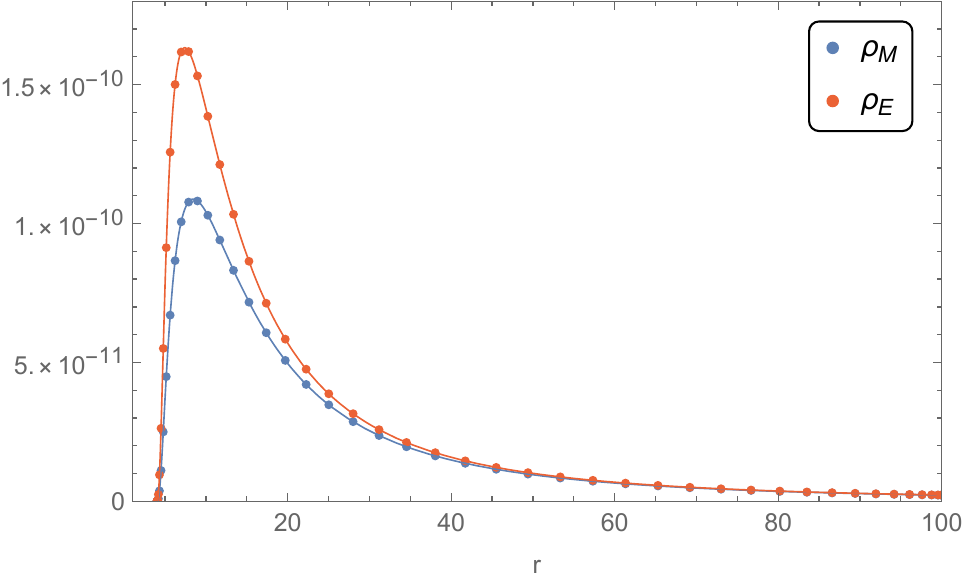}
	\caption{The comparison between the mass density $\rho_{M}$ (blue) and the energy density $\rho_{E}$ (red) of the DM spike. At the peak, the maximum of $\rho_{E}$ is approximately $1.50$ times the maximum of $\rho_{M}$.}
	\label{fig:rho vs Ttt}
\end{figure}
This section presents our results for both the energy-momentum tensor of the DM spike and the corresponding BH solution influenced by it. As in previous sections, the BH mass is fixed at \( M_{\text{BH}} = 1 \) for all calculations.

We begin with the radial distributions of the mass density (upper panel) and the components of \( T_{\mu}^{\nu} \) (lower panel) in Fig.~\ref{fig:rho T vs r}. The dotted and dashed lines in both panels mark the radius of the BH horizon and the marginally bound orbit, respectively. The upper panel compares the initial Hernquist profile \( \rho^{(H)}_{M} \) (black) with the mass density \( \rho_{M} \) (blue) after the adiabatic growth of the BH. The formation of a dense spike in \( \rho_{M} \), orders of magnitude higher than the initial profile, successfully reproduces the findings of \cite{Sadeghian:2013bga}.
\begin{figure*}[t!]
	\begin{center}
		\mbox{ 
			\includegraphics[width = 0.84\textwidth]{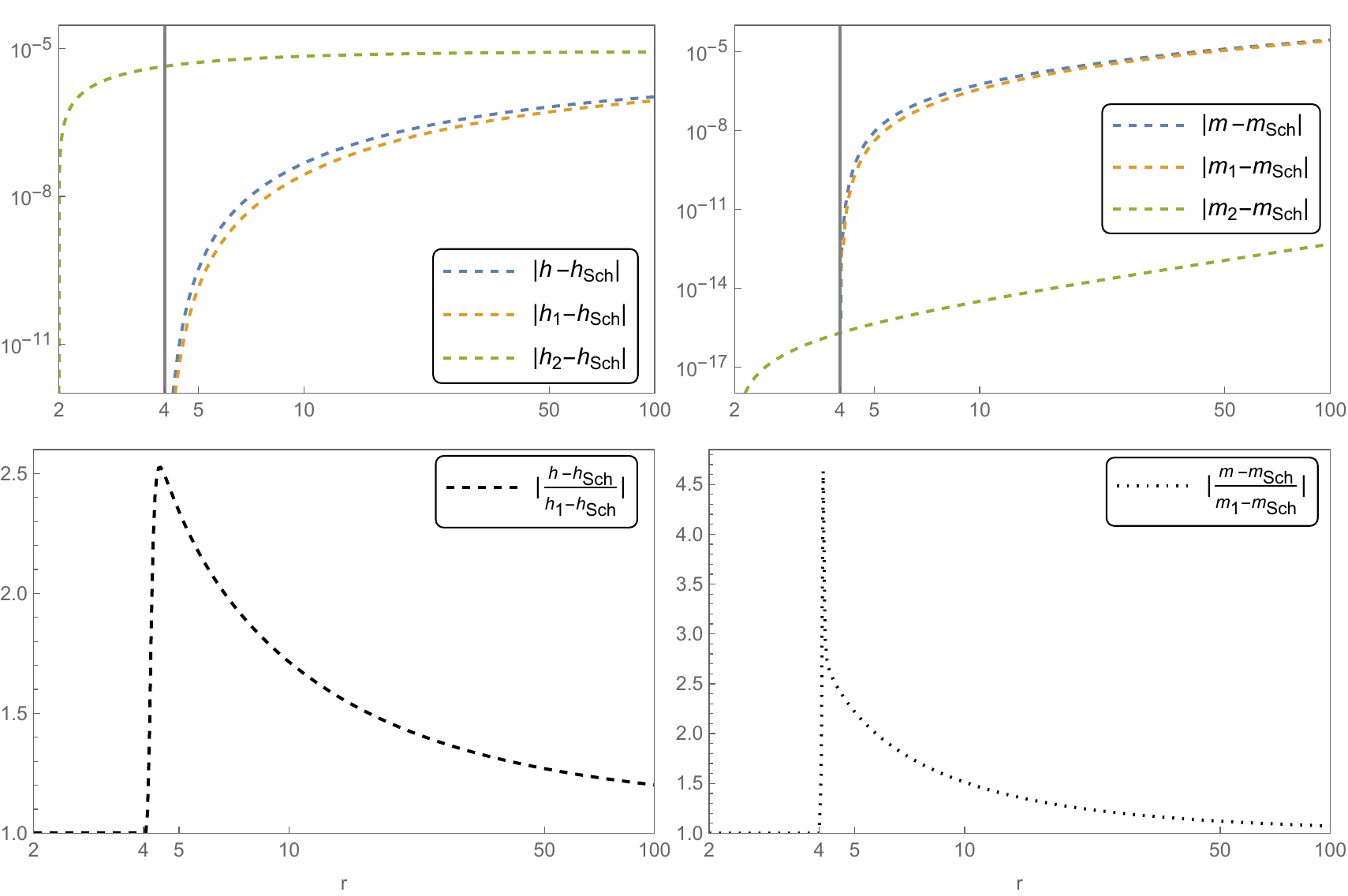}
		}
		\vspace*{-0.5cm}
	\end{center}
	\caption{The deviation of the numerical solutions ($h$,$h_1$,$h_2$) (upper left) and ($m$,$m_1$,$m_2$) (upper right) from their Schwarzschild counterparts \( g_{\textrm{Sch}}\) and \( m_{\textrm{Sch}}\), and the ratio of deviations of our resulting metric to those of \( g_{(1)}\) and \( m_{(1)} \)in the bottom two panels. The subscript ''1'' of ($h_1$, $m_1$) represents that the results is obtained by an alternative approach from \cite{Chakraborty:2024gcr}, and the subscript ''2'' of ($h_2$, $m_2$) corresponding to the alternative approach from \cite{Figueiredo:2023gas}. The gray thin lines in upper panels show the radius of marginally bound orbit for the particles on the Schwarzschild BH, indicating that the DM spike abruptly vanishes at this radius. 
	}
	\label{fig:delta and ratio}
\end{figure*} 
In the lower panel, the components of the energy-momentum tensor reveal a hierarchy in their peak magnitudes within the spike; the energy density \( \rho_E \) surpasses the radial pressure by two orders of magnitude. This finding offers some justification for the neglect of radial pressure in studies like \cite{Cardoso:2021wlq,Chakraborty:2024gcr}. However, it is crucial to note that this result is derived under the specific assumption that all particles at a given point follow circular geodesics \cite{Comer:1993a,Comer:1993rx}. 
We emphasize that the radial component \( T_{r}^{r} \) is nonvanishing. Beyond the spike region, the radial pressure \( p_r \) and tangential pressure \( p_t \) become comparable in magnitude, indicating that \( p_r \) cannot be neglected, at least in the far zone. In fact, \( T_{r}^{r} \) can vanish only if the square root term in Eq.~(\ref{eq:Trr}) is zero. For the integral in Eq.~(\ref{eq:Trr}) at a fixed \( r \), generic elliptical geodesics with nonzero square roots (except at \( r=4 \) or \( \infty \)) contribute, leading to a nonzero \( T_{r}^{r} \). 

To validate our solution, we check for potential violations of the energy conditions by the derived energy-momentum tensor. The conditions are defined as follows. The Null Energy Condition requires  $\rho_E+p_i\geq0$, the Weak Energy Condition requires $\rho_E\geq0$ and $\rho_E+p_i\geq0$, the Strong Energy Condition requires $\rho_E+p_i\geq0$ and $\rho_E+\sum_i p_i\geq0$, and the Dominant Energy Condition requires $\rho_E\geq|p_i|$. Here, $p_i$ refers to any spatial component of the tensor given in Eq. (\ref{eq:EnergyMomentumTensor}). As is clear from Fig.~\ref{fig:rho T vs r}, all these energy conditions are respected in the region occupied by DM. Notably, as shown in \cite{Datta:2023zmd}, when the DM spike is modeled as either an anisotropic fluid with vanishing radial pressure or an isotropic fluid, and the mass density is constructed by multiplying the halo profile with a cutoff factor (e.g. \cite{Cardoso:2021wlq}), the dominant energy condition is typically violated unless the cutoff is chosen appropriately. In contrast, the DM spike in this work is modeled in a self-consistent manner, so the assumptions adopted in \cite{Datta:2023zmd} do not apply here.

The comparison between the mass density $\rho_M$ and the energy density $\rho_E$ is shown in Fig.~\ref{fig:rho vs Ttt}, highlighting the limitation of using $\rho_M$ as a substitute for $\rho_E$ when determining the metric under the influence of DM. In the vicinity of the DM spike, particles influenced by the BH exhibit both high density and non-negligible kinetic energy. Consequently, a substantial portion of the energy sourcing the gravitational field is carried in this kinetic form. Figure~\ref{fig:rho vs Ttt} shows that the peak value of $\rho_E$ is approximately 1.50 times that of $\rho_M$, indicating that about 50\% of the energy density in the spike manifests as kinetic energy of the DM particles. Beyond the spike, the difference between the two densities diminishes rapidly, as the maximum velocity of bound particles in the far region becomes increasingly small (with the specific relativistic energy approaching unity). This behavior is consistent with the analysis of the radial potential Eq.~(\ref{eq:effective potential}) in Sec.~\ref{section DM spike}. 

We present the deviation (the blue dashed line) of numerical metric functions \( h(r) \) and \( m(r) \), from their Schwarzschild counterparts  (\( h_{\textrm{Sch}}, \ m_{\textrm{Sch}} \)) around the spike (\( r_{H} \leq r \leq 10^2 \)) in the upper panels of Fig.~\ref{fig:delta and ratio}. The upper left panel shows that the modified metric function \( h(r) \) exhibits a slight but yet distinct deviation (the blue dashed line) from the Schwarzschild case, on the order of \( 10^{-8} \)-\( 10^{-6} \) around the middle region ($10<r<10^2$). Notably, the magnitude of this deviation exceeds that of the energy-momentum tensor itself. The deviation of mass function \( m(r) \), shown by the blue dashed line in the upper right panel, reveals a more pronounced difference from the Schwarzschild solution, with deviations of order \( 10^{-5} \) around the middle region. 

The deviations associated with alternative approaches are also shown in the upper panel of the Fig.~\ref{fig:delta and ratio}. The yellow dashed line corresponds to the approach of Ref.~\cite{Chakraborty:2024gcr}, which employs the resst-mass density of DM spike generated by the adiabatic growth of the BH and assumes circular particle orbits ($p_r=0$). The green dashed line represents the approach of Ref.~\cite{Figueiredo:2023gas}, where the energy density of the DM spike is obtained by multiplying the original Hernquist profile (\ref{eq:Hernquist profile}) with an additional factor ($1-2M_{BH}/r$) and under the assumption that the anisotropic fluid has a vanishing radial pressure. These two approaches are calculated on the full relativistic equations. The key features and differences among these approaches are summarized in Table~\ref{table1}. 

The second alternative approach of Ref.~\cite{Figueiredo:2023gas} (the green dashed line) exhibits behavior that is qualitatively distinct from the other two methods. In this case, the deviation in $h_2$ is significantly larger, whereas the deviation in $m_2$ is considerably smaller when compared to the other approaches. This feature originates from its treatment of the effective energy density component [$T_t{}^t =- (1-2M_{BH}/r)\rho^{(H)}_{M} (r)$], which allows DM particles to remain supported within the radius of the marginally bound orbit ($r=4$). When this prescription is formally extended to radii smaller than the event horizon, the corresponding effective energy density becomes negative, leading to a physically pathological modification of the BH spacetime inside the horizon. As a consequence, the green dashed line coincides with the Schwarzschild solution only precisely at the horizon radius, while deviating from it both outside and (though not shown) inside the horizon. By contrast, the blue and yellow dashed lines recover the Schwarzschild solution exactly for $r \leq 4$.

\begin{table}[htbp]
\centering
\caption{Comparison of different approaches adopted in this work and in the literature.}
\label{tab:comparison}
\begin{tabular}{lccc}
\hline\hline
 & Adiabatic growth  & Full particle  & Full relativistic \\
 & of a BH & orbits &  equations \\
\hline
this work           & \checkmark & \checkmark & $\times$ \\
$(h,m)$         & &  &  \\
Ref.~\cite{Chakraborty:2024gcr}  & \checkmark & $\times$     & \checkmark \\
$(h_1,m_1)$       & &  &  \\
Ref.~\cite{Figueiredo:2023gas}  & $\times$     & $\times$     & \checkmark \\
$(h_2,m_2)$        & &  &  \\
\hline\hline
\end{tabular}
\label{table1}
\end{table}

Although these deviations are small due to the use of Milky Way parameters, our resulting metric exhibits a stronger DM effect than that (\( h_{1},  m_{1} \)), which only utilizes the mass density of the DM spike. The ratios of the deviation between the two comparable approaches, defined as \((h-h_{\textrm{Sch}})/(h_{1}-h_{\textrm{Sch}}) \) and \((m-m_{\textrm{Sch}})/(m_{1}-m_{\textrm{Sch}})\), are depicted by the black dashed line and black dotted line in bottom panels of Fig.~\ref{fig:delta and ratio}, respectively. Particularly, the deviations in our results both exceed those of the alternative method by a factor of 2.5 (4.6) for $h$ ($m$).  Two key differences distinguish our method. First, we use the full energy density \( \rho_E \) Eq.~(\ref{eq:Ttt}) rather than the rest-mass density \( \rho_M \) Eq.~(\ref{eq:jt}). Second, we consistently include the nonvanishing radial pressure \( p_r \) Eq.~(\ref{eq:Trr}), as derived from the adiabatic growth model.  This enhancement is primarily attributed to the inclusion of particle kinetic energy. Consequently, our model captures not only the enhanced density of DM particles near the BH but also the gravitational influence of their significant  orbital motion.

\section{discussion}
\label{discuss}
In this work, we have studied the energy-momentum tensor of a DM spike formed through the adiabatic growth of a BH within a galactic halo, and investigated its backreaction on the spacetime geometry. Within the Einstein cluster framework, by adopting the Hernquist density profile as a representative example of DM halo and employing parameters consistent with those of the Milky Way, we derived a statistical formulation of the full tensor, explicitly including the kinetic contribution to the energy density and the anisotropic pressure arising from noncircular orbits. Our results show that the kinetic term enhances the total energy density near the spike by approximately 50\% compared with the rest-mass component, while a small but nonvanishing radial pressure introduces mild anisotropy in the stress tensor. We investigate the backreaction of the DM spike on spacetime, beginning with an analysis within a perturbative framework. Using the energy-momentum tensor of DM spike as a fixed source in the equations for a combined metric, we numerically obtained a static, spherically symmetric solution modified by the DM distribution. The resulting spacetime exhibits a slight but measurable deviation from the Schwarzschild solution --- roughly a factor of two larger than that found in previous treatments considering only the mass density. Furthermore, the derived energy-momentum tensor satisfies all standard energy conditions, confirming the internal consistency of the model and the physical viability of our relativistic description of the DM spike. 

Compared with previous treatments that modeled the DM spike as a static or purely circular-orbit fluid \cite{Chakraborty:2024gcr}, our approach provides a more realistic and dynamical description of the system. The inclusion of kinetic energy and radial pressure naturally leads to stronger deviations of the metric than in earlier simplified analyses, while remaining physically justified since realistic DM spikes around supermassive BHs are expected to be extremely dilute and to generate only weak gravitational fields. This confirms that the semiconsistent iterative procedure employed here captures the dominant physical effects of the DM backreaction while maintaining computational tractability.

Although our model is semiconsistent --- the metric and distribution function are obtained iteratively rather than solved simultaneously --- it provides a physically grounded step toward a fully self-consistent relativistic description of BHs embedded in DM halos. Achieving complete self-consistency would require recalculating the DM distribution in a numerically determined metric throughout the adiabatic growth process, which poses a significant computational challenge. While the present analysis adopts a Hernquist profile for the initial halo, the method can be readily generalized to other realistic DM halo models such as NFW \cite{Navarro:1996gj}, Einasto \cite{Retana-Montenegro:2012dbd}, or cored isothermal profiles \cite{Weber:2009pt}.\footnote{The latter, however, should be regarded as effective parametrizations rather than equilibrium solutions of collisionless systems, since core-like structures generally require additional physics such as self-interactions or baryonic feedback.}  Extending this framework to include rotating (Kerr) BHs (a completed work that employs a post-Newtonian setup is presented in \cite{Mitra:2025tag}), nonadiabatic evolution, more general DM distributions, and dynamical analyses related to gravitational wave emission will be the focus of future work.

\begin{acknowledgments}
	The work is in part supported by NSFC Grant No. 12205104 and the startup funding of South China University of Technology.
\end{acknowledgments}

\end{document}